\documentclass[preprint,12pt]{elsarticle}
\usepackage[margin=2.5cm]{geometry}
\usepackage{siunitx,booktabs}
\usepackage{threeparttable}
\usepackage[export]{adjustbox}[2011/08/13]  
\usepackage[labelfont=bf,justification=raggedright,singlelinecheck=false]{caption}
\makeatletter
\def\ps@pprintTitle{%
 \let\@oddhead\@empty
 \let\@evenhead\@empty
 \def\@oddfoot{}%
 \let\@evenfoot\@oddfoot}
\makeatother

\usepackage{graphicx}
\usepackage{amssymb}

\usepackage{lineno}

\begin{document}

\begin{frontmatter}


\title{The Wisdom of Polarized Crowds}



\author[a,*]{Feng Shi}
\author[b,*]{Misha Teplitskiy\corref{cor1}} 
\author[c,d]{Eamon Duede}
\author[d,e]{James A. Evans\corref{cor1}}

\address[a]{Odum Institute for Research in Social Science, University of North Carolina at Chapel Hill}
\address[b]{Laboratory for Innovation Science, Harvard University}
\address[c]{Committee on the Conceptual and Historical Studies of Science, University of Chicago}
\address[d]{Knowledge Lab, University of Chicago}
\address[e]{Department of Sociology, University of Chicago}

\fntext[*]{FS and MT contributed equally to this work.}
\cortext[cor1]{To whom correspondence should be addressed: mteplitskiy@fas.harvard.edu; jevans@uchicago.edu}

\begin{abstract}
As political polarization in the United States continues to rise, the question of whether polarized individuals can fruitfully cooperate becomes pressing. Although diversity of individual perspectives typically leads to superior team performance on complex tasks, strong political perspectives have been associated with conflict, misinformation and a reluctance to engage with people and perspectives beyond one's echo chamber. It is unclear whether self-selected teams of politically diverse individuals will create higher or lower quality outcomes. In this paper, we explore the effect of team political composition on performance through analysis of millions of edits to Wikipedia's Political, Social Issues, and Science articles. We measure editors' political alignments by their contributions to conservative versus liberal articles. A survey of editors validates that those who primarily edit liberal articles identify more strongly with the Democratic party and those who edit conservative ones with the Republican party. Our analysis then reveals that polarized teams---those consisting of a balanced set of politically diverse editors---create articles of higher quality than politically homogeneous teams. The effect appears most strongly in Wikipedia's Political articles, but is also observed in Social Issues and even Science articles. Analysis of article ``talk pages'' reveals that politically polarized teams engage in longer, more constructive, competitive, and substantively focused but linguistically diverse debates than political moderates. More intense use of Wikipedia policies by politically diverse teams suggests institutional design principles to help unleash the power of politically polarized teams.
\end{abstract}


\end{frontmatter}


\section{Introduction}
\label{S:1}

Recent political events, including the 2016 presidential election, have underscored growing political divisions in American society. Political speech has become markedly more polarized in recent years \cite{Gentzkow2016-wh}, tracing a growing divergence between platforms of the major political parties \cite{Fiorina2008-xr} and leading to a state of political hyper-partisanship \cite{Campbell2016-oa}. Yet the effects of political difference are not confined to the domain of politics alone. A growing literature documents how individual political alignments shape personal consumption of ostensibly non-political products, news, cultural and scientific information \cite{DellaPosta2015-hw,Gauchat2012-od,Sarewitz2004-vb,Zhao2011-vl,Shi2017-sq}. This literature has converged on an alarming narrative: despite early promise of the world-wide-web to democratize access to diverse information \cite{benkler2006wealth}, increased media choice and social networking platforms have led to the converse. Collaborative filtering allows individuals to passively enter ``echo chambers'' that limit the variety of information they observe and trust \cite{Bakshy2015-fc,Pariser2011-ag,Del_Vicario2016-lh}. These can degrade the quality of individual decisions, including those that undergird basic democratic institutions \cite{sunstein2017republic,Mutz2006-ci,Bishop2009-mu}. Psychological mechanisms such as motivated reasoning \cite{Taber2006-hx,miller2016conspiracy} and a tendency to discount identity-incongruent opinions \cite{Mutz2006-ci,Kahan2017-fa} stimulate and reinforce polarizing information. Opposing social identities can foment conflict and even make communication counter-productive \cite{Hart2012-xn}. 

Nevertheless, a large literature documents the largely positive effect that social differences can exert on the collaborative production of information, goods and services \cite{Joshi2009-ot,Page2008-gu}. Research demonstrates that individuals from socially distinct groups embody diverse cognitive resources and perspectives that, when cooperatively combined in complex or creative tasks produce ideas, solutions, and designs that outperform those from homogeneous groups \cite{Mannix2005-sn,Hong2004-xg,Woolley2010-mc,Nielsen2012reinventing}. Collaborations between inventors from distinct social groups result in more creative patents \cite{Fleming2007-kk}, scientific teams representing distinct disciplines produce more highly cited papers \cite{Wuchty2007-dp}, and gender diversity broadens the questions scientists ask \cite{Nielsen2017-ea}. 

The effect of political diversity on the collective production of knowledge, however, remains unclear. Insights from cognitive diversity research suggest that political diversity, like other forms of diversity, should positively impact the quality of group production. Literature on echo chambers suggests that political diversity may hamper productive cooperation, however, as partisans perceive information held by opponents as not simply different, but wrong. In short, political diversity should increase access to fresh perspectives and information but may also undermine the quality of discourse and engagement required to enjoy the performance benefits typically obtained by diverse groups.

In order to assess the effect of political diversity on team performance, we studied the effect of political polarization on the performance of approximately four hundred thousand 
online teams. Specifically, we focused on teams of Wikipedia editors who worked on English-language articles in three large domains: Politics, Social Issues, and Science. 
  
\section*{Data and Methods}
Using edit histories, we measured the political alignments of millions of Wikipedia editors by the relative amount of content they contributed to conservative versus liberal political articles. We validated this measure by surveying a random sample of Wikipedia editors for whom we had calculated the index. We then used a machine learning algorithm developed by Wikimedia's internal researchers to measure the quality of Wikipedia articles  \cite{Halfaker2017-cx}. We finally related article quality to the political diversity of teams, and, to gain insight regarding the mechanisms of collaboration among polarized teams, we computationally explored characteristics of article ``talk pages'' where the work of editing and debate occurs. 

\subsection*{Data collection}
We extracted data from the complete English Wikipedia database dump on 12/01/2016. Data includes all edits made to all English Wikipedia pages since its start until 12/01/2016. Within this dump, we focused on three sets of articles: politics (20,947 pages), social issues (162,085 pages) and science (49,530 pages), which represent approximately 5\% of all English Wikipedia articles. Summary statistics of the three corpora may be found in Table \ref{tab:stats}. Users' total numbers of edits ever made to Wikipedia were collected through Wikipedia's online API.\footnote{http://en.wikipedia.org/w/api.php}.

The corpus of Political pages consists of two sub-corpora, Liberal and Conservative pages. The Liberal sub-corpus consists of all pages categorized under the ``American liberalism'' category and its subcategories. For instance, the page ``New Deal coalition'' is directly under the ``American liberalism'' category, while ``The New Republic'' is located under the sub-category ``American liberalism > Modern liberal American magazines''. The Conservative sub-corpus was collected in a similar fashion starting with the ``American conservatism'' page. For instance, ``American Conservatism'' links to ``Economic liberalism,'' which links to ``Market economy,'' and all three pages are in the ``Conservative'' sub-corpus. Pages appearing in both corpora were removed. 

Titles of Social Issues pages were collected starting from the page ``Category:Social issues" \footnote{https://en.wikipedia.org/wiki/Category:Social\_issues}. We collected all pages and subcategories linked from the page; repeating this process in every subcategory of Social Issues, stopping 4 levels down from the root. Social Issues include articles relating to human welfare and justice, including ``Homelessness," ``Teenage pregnancy," and ``Social services." These pages tend to be relatively controversial and politically salient. Titles of science pages were collected similarly, following the category structure of scientific disciplines in Wikipedia, starting from the page ``Category:Scientific disciplines" \footnote{https://en.wikipedia.org/wiki/Category:Scientific\_disciplines} and following the iterative procedure pursued for Social Issues pages.

\subsection*{Survey of Wikipedia editors}
To validate our statistical measure of political alignments, we surveyed a random sample of editors for whom we had estimated alignment scores. We worked directly with the Wikipedia community and Wikimedia staff to carry out the survey, including the development of a research page on the Wikimedia ``Meta-Wiki'' site and direct engagement with those expressing concerns therein\footnote{https://meta.wikimedia.org/wiki/Research:Wikipedia\_\%2B\_Politics}. The arrived-upon process required a single member of our team (E.D.) to personally post the survey link on each editor's page along with an explanation. The number of solicitations we could make per day (and their total number) was capped. In the end, we were able to post 500 solicitations\footnote{The survey may be viewed here: https://uchicago.co1.qualtrics.com/jfe/form/SV\_eXOHLbXwbpfYC1f} and received 118 responses. The survey was approved by the University of Chicago's Institutional Review Board (IRB17-0679). More information, including response rates by (computationally measured) political alignment may be found in Appendix.

\subsection*{Measurement}
For each user, we used total size (in bytes) of contributions she made to liberal (blue) versus conservative (red) articles to infer her political alignment. Specifically, we model the total bytes she contributed to red articles ($X$) as a random variable satisfying a binomial distribution $X\sim Binomial(K,p)$, where $K$ is the total number of bytes contributed to political articles (red or blue) and $p$ is the probability of contributing to red articles. This probability $p$ represents our measure of \textbf{political alignment} for this editor, after rescaling it to the range -1 (most liberal) to +1 (most conservative).  The parameter $p$ is an unknown quantity to be estimated from observations $X$ and $K$. We estimated it through a conservative, Bayesian framework described in Appendix. The quantity of primary interest is the variance of alignments among a group of editors, which quantifies the spread of editors across the liberal-conservative spectrum. We used the variance of political alignments as the measure for polarization of any group of editors. Previous research has found that this measure most directly captures the cognitive diversity of a group along a particular demographic dimension \cite{Bell2010-uc}.

\section*{Results}

\subsection*{Editors' political alignments}
We measure editors' alignments by the fraction of bytes they contribute to ``Conservative'' versus ``Liberal'' articles on the English-language Wikipedia, with a Bayesian framework to account for random edits. The corpus of conservative articles consists of all articles categorized under ``Conservatism in the United States\footnote{https://en.wikipedia.org/wiki/Category:Conservatism\_in\_the\_United\_States.},'' and similarly for ``Liberalism in the United States\footnote{https://en.wikipedia.org/wiki/Category:Liberalism\_in\_the\_United\_States.}.'' This procedure scores editors as politically neutral  ($\approx 0$) if they contribute equally to both sets of articles or little to either set, and closer to -1 or +1 the more exclusively they contribute to liberal or conservative articles, respectively. 

\begin{figure*}
\centering
\includegraphics[width=1.01\linewidth]{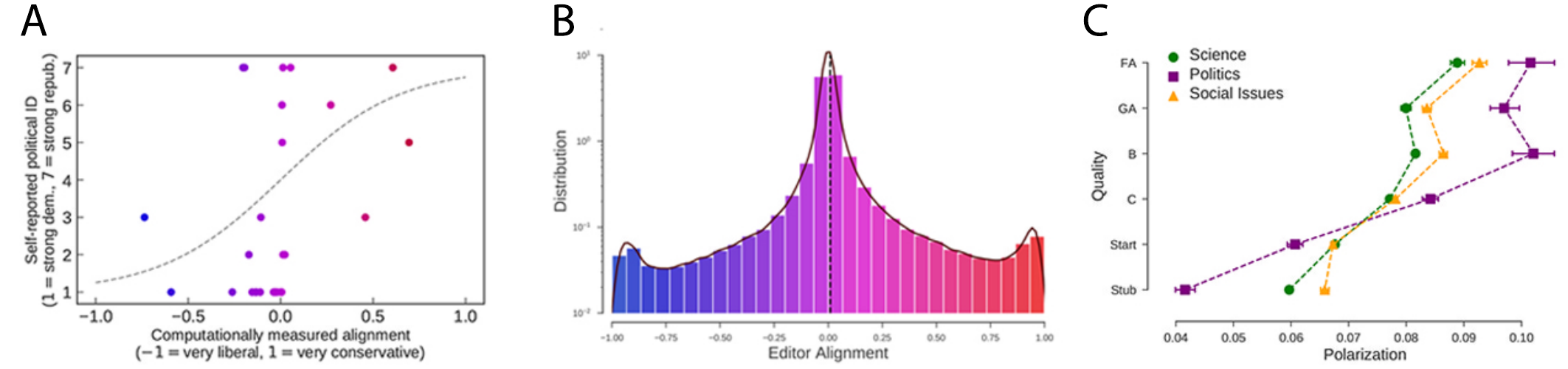}
\caption{\textbf{A}. Scatter plot of implied political alignment (-1=most liberal, +1=most conservative) and political identification (1=Strong democrat, 4=Independent, 7=Strong republican) reported by US-based survey respondents. Respondents identifying as ``Independent" were excluded from analysis, and 7 responses of ``Other" were recoded to either ``3=Independent, Near Democrat" or ``5=Independent, Near Republican" (see Appendix for details). Dotted line is the best-fitting Logistic sigmoid, and its curvature suggests that even those at the boundary of our editing measure tend to ``switch" between Republican and Democratic identification. Pearson correlation coefficient between the two measures is $\rho=0.35$ ($n=28, p=0.036$, 1-tailed \textit{t}-test). \textbf{B}. Distribution of editors' computationally measured political alignments. \textbf{C}.  Article quality (Stub=lowest, FA=highest) by average team polarization for Politics (purple), Social Issues (orange) and Science (green) articles. Bands around each mean denote its \%95 confidence interval. Lines are best linear fits to the points in the plot.}
\label{fig:alignment_measure}
\end{figure*}

118 responses from a survey targeted at randomly chosen editors of science and political articles validate our computational measure of political alignment (Figure \ref{fig:alignment_measure} A). Respondents' self-reported political party identification correlates at roughly 0.35 with our computational measure of conservative-liberal alignment, and validates our use of editing history as a (noisy) behavioral indicator of political preferences. 

With inferred political alignments, we observe that Wikipedia editors display a wide distribution of political alignments (Figure \ref{fig:alignment_measure} B). The peak at the center of the distribution comports with our observation that a large number of people only contributed minor edits to Wikipedia, such as correcting a typo. There are also two lower but significant peaks at the tails of the distribution, which identify editors who contribute substantial content to either liberal or conservative articles and suggest substantial polarization on Wikipedia. The variance of alignments across all editors of political articles is 0.04, significantly higher than random (See Appendix for details on random simulations). We then measure the \textit{polarization} of any given group of editors by the variance of their alignment scores.

\begin{figure}
\centering
\includegraphics[width=0.9\linewidth]{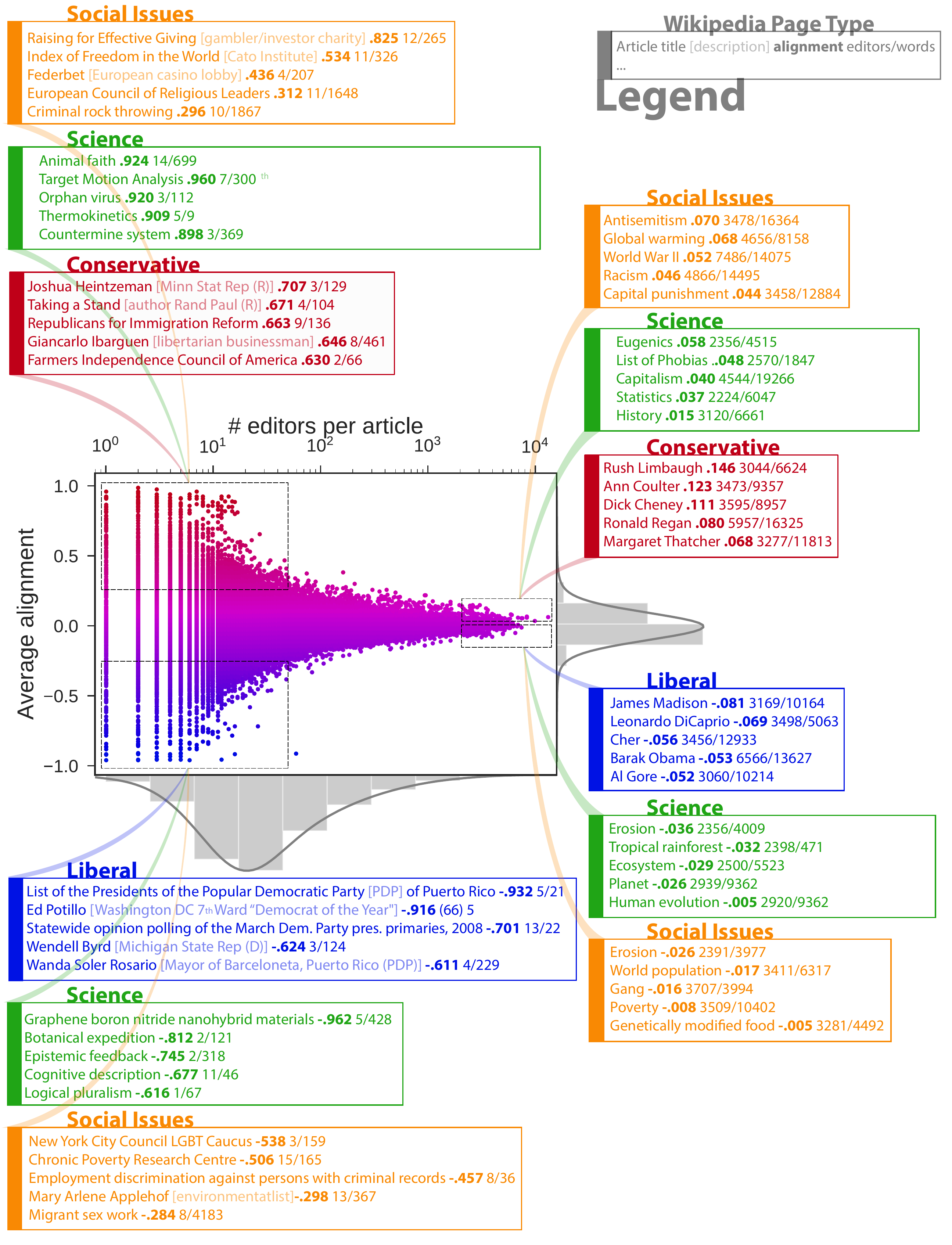}%
\caption{Scatter plot of each article's average editor alignment by number of editors. Average political alignment converges to 0 as the number of editors increases, demonstrating the Linus Effect. Histograms on x and y axes reveal the density of articles at each level of editorial attention and average political alignment, respectively. Call-out boxes list five of the most ``liberal'' and ``conservative'' pages for articles receiving the most and least editorial attention, featuring article title followed by an optional description, mean political alignment, number of editors, and article length in bytes. These examples illustrate meaningful association between right and left political preference of Wikipedia editors and the pages they edit (e.g., ``capitalism'' and ``history'' vs. ``planet'' and ``human evolution''.}
\label{fig:article_alignment}
\end{figure}

As the number of editors for an article increases, their average political alignment converges to 0 (Figure \ref{fig:article_alignment}). This phenomenon is sometimes referred to as Linus' law -- ``with enough eyeballs, all bugs are shallow.''  In our case, articles attracting more attention tend to have more balanced engagement from editors along the conservative-liberal spectrum. This finding replicates those reported by Greenstein and Zhu \cite{Greenstein2012-xj,Greenstein2016-ei} in their studies of bias in Wikipedia's US political coverage, showing that increased editor interaction reduced individual biases and yielded greater content neutrality. 

\subsection*{Effects on Quality}
We measure the quality of articles using a machine learning model developed by Wikimedia research staff and trained on features of article content alone -- no features of the editors were used to train the model. The 6-category quality scale for Wikipedia articles ranges from ``Featured article" (highest quality) to ``Stub'' (lowest quality). Figure \ref{fig:alignment_measure} C plots the relationship between average team polarization (i.e., variance of alignments) and quality for Political, Social Issues, and Science articles. 

In all three corpora -- Political, Social Issues, and Science articles -- higher polarization is associated with higher quality. To establish this relationship statistically, we estimated an ordinal logistic regression model at the article level with article quality as outcome and polarization as main independent variable. We added the absolute value of mean team alignment to account for the possibility that article quality is related to the deviation of political alignment from neutral (0) in either direction. Additionally, we added controls for article and editor features that may plausibly confound the relationship between polarization and quality. These include length, number of edits, and number of editors for each article, and average editing experience for the editors (see Data and Methods for details).

Regression results are provided in Table \ref{tab:quality}. As expected, number of edits, length of article, and number of editors significantly predict article quality. The coefficient for the |alignment| term suggests that quality decreases when editors are biased, on average, in either direction. Most critical is that polarization, the variance of political alignments, is positively and substantially associated with quality: a 1-unit increase in polarization multiplies the odds of moving from lower- to higher-quality categories by a factor of 18.57 for Political articles, 2.06 for Social Issues articles and 1.90 for Science articles. 

\begin{table}
\centering
\small
\begin{threeparttable}
\caption{Odds ratios from ordinal logistic regression models predicting article quality}
\label{tab:quality}
\begin{tabular}{r *{3}{S[table-format=1.3]}}
& \multicolumn{3}{c}{\textbf{Dependent variable: article quality}} \\
\textbf{Independent variable} & \textit{Politics} & \textit{Social issues} & \textit{Science} \\
\midrule 
polarization                  & 18.88 ***                        & 2.06 *** & 1.79 **                    \\
$|$ alignment $|$                 & 0.30 ***                        & 0.49 *** & 0.65 **                      \\
editing experience		              & 1.05 *                          & 1.06 *** & 1.01                       \\
number of editors                    & 0.41 ***                        & 0.51 *** & 0.56 ***                      \\
article length                        & 33.55 ***                        & 47.83 *** & 56.54 ***                    \\
number of edits                      & 3.26 ***                        & 1.71 *** & 1.69 ***                      \\
\multicolumn{3}{l}{\textit{}}    \\
\textit{N}                    & \multicolumn{1}{c}{12,570}      &  \multicolumn{1}{c}{161,070}  & \multicolumn{1}{c}{49,995}\\
\bottomrule
\end{tabular}
\begin{tablenotes}
\small
\item
\textit{Note}: *, **, *** denote statistical significance levels of 0.1, 0.01 and 0.001, respectively. The columns present odds ratios estimated on Political, Social issues and Science articles, separately.
\end{tablenotes}
\end{threeparttable}
\end{table}

\subsection*{Mechanisms of Polarized Collaboration}
To explore mechanisms by which politically polarized teams outperform homogeneous teams, we examine Wikipedia `talk pages'. Each Wikipedia article has an associated talk page where `backstage' knowledge assemblage occurs. Here, editors debate proposed additions and deletions, identify shortcomings, and attempt to persuade their fellow editors regarding content for the public facing, `frontstage', Wikipedia article \cite{Viegas2007-et}. Using text from these talk pages, we examine relationships between political polarization and the following aspects of debate: (1) debate intensity, (2) information diversity, and (3) use of Wikipedia institutions---policies and guidelines---to discipline discussion. We investigate pairwise correlations between polarization and these debate mechanisms, then we estimate regression models to test the effect of polarization on these mechanisms separately, and finally, assemble them into a structural equation model that allows us to identify their relative influence on article quality. All statistical analyses yield consistent results regarding mechanisms of collaboration, as discussed below and detailed in the Appendix. 

\begin{figure}
\centering
\includegraphics[width=0.6\linewidth]{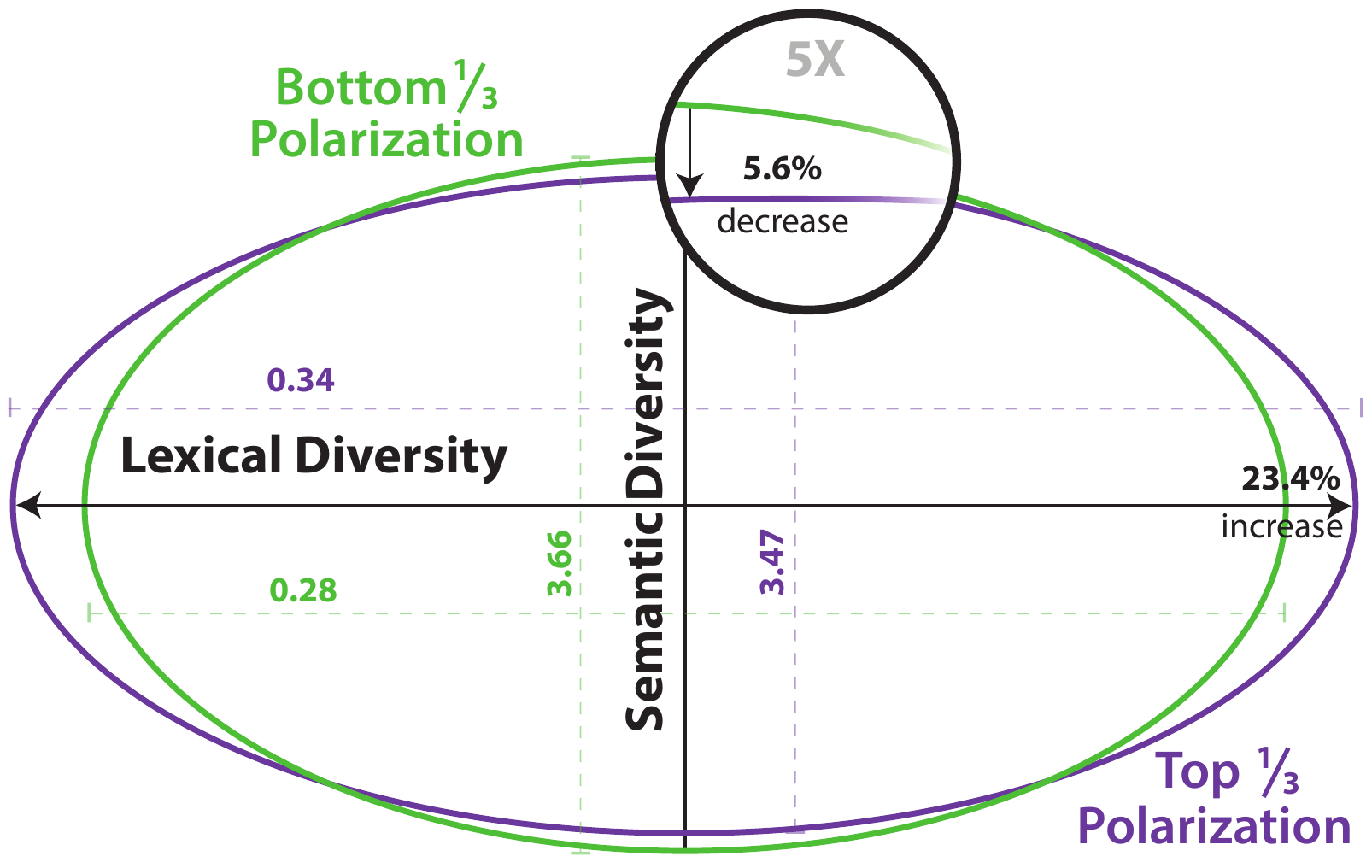}%
\caption{Illustration of the shift in ``talk page'' debate activity between teams in the bottom and top thirds of the political polarization distribution. Compared with the least polarized teams, the most polarized teams semantically contract by 5.6\% and lexically expand by 23.4\%: they talk more about less, focusing on core politically-contested subjects, but framing them in distinctive ways.}
\label{fig:diversity}
\end{figure}

Studies of team diversity and performance argue that information diversity is the key feature distinguishing diverse from homogeneous teams. Nevertheless, this is almost never measured directly, particularly in non-laboratory settings. Here, we decompose ``information diversity'' into two distinct dimensions: \textit{lexical} and \textit{semantic} diversity. \textit{Semantic diversity} traces distinct meanings or issues discussed in a talk page, while \textit{lexical diversity} captures the number of ways in which editors discuss them. We expect that political polarization will focus debate on a few contested, politically relevant topics, but frame them in multiple ways, yielding lower semantic diversity and higher lexical diversity. We measure lexical diversity of each talk page as a function of its distinct and distinguishing words. We measure the semantic diversity of a page as a function of the dispersion of words on that page in a latent semantic space defined by all Wikipedia articles, such that higher semantic diversity indicates more Wikipedia topics were debated. (See Data and Methods for details on the two diversity measures.) We find that high polarization narrows debate by reducing talk page semantic diversity, but generates alternative framings traced by greater lexical diversity, as illustrated in Figure \ref{fig:diversity}.  

Diverse information should be more difficult to integrate, particularly if contested. We measure two core aspects of debate intensity including \textit{volume} and \textit{temperature}. Following previous research that found talk page length associated with article quality \cite{Wilkinson2007-us,Kittur2008-gb}, we measure debate volume as a function of talk page length and distinct edits. Polarized teams may attempt to integrate more diverse information, requiring more talk, which yields greater article quality.  Integrating diverse perspectives on contested and value-laden topics could be acrimonious, but a balance of liberals and conservatives could lower the \textit{temperature} of potentially volatile collaborations, following research that links competitive imbalance to emotional aggression and violence \cite{collins2009violence}. We measure debate temperature using the \texttt{Detox} tool, developed by Wikimedia to identify harassment in the Wiki community. \texttt{Detox} detects toxic comments using a sophisticated machine learning classifier \cite{wulczyn2017ex}, which we apply to all talk page edits. We find that polarized teams generate a larger volume of debate and their balance of political perspectives reduces flare-ups in debate temperature.

Finally, we explore the self-governance of contested knowledge through use of Wikipedia policies and guidelines. Policies and guidelines are invoked so frequently that they have a standard nomenclature \footnote{https://en.wikipedia.org/wiki/Wikipedia:Shortcut\_directory.}. For example, an editor who believes that part of an article is biased may invoke ``NPOV'' (the ``Neutral Point of View'' policy) in the article's talk page. Wikipedia also relies on a collection of less binding guidelines that refer to desired qualities of Wikipedia pages and the editorial process. These include that articles should cite sources (``CITE'') and avoid and/or disclose any conflicts of interest (``COI''). We expect editors within polarized teams to encounter differences not easily resolved and, when debate fails, to discipline or challenge collaborators by invoking Wikipedia's policies and guidelines. Indeed, the numbers of policy and guideline mentions are found to increase with polarization. When disaggregated, we find that ``NPOV'' (Neutral point of view) and ``OR'' or ``NOR'' (No original research) are the most frequently cited policies, each significantly correlated with polarization.

Correlations between all modeled variables are presented in Figure S2 and are consistent with the regressions and structural equation model described below (also see Appendix). We also note interesting associations \textit{between} talk page measures, suggesting micro-mechanisms of conflict and coordination, such as the negative correlation between debate temperature and volume. This is relevant to the growing literature about online ``trolling'' behavior \cite{cheng2014community, cheng2015antisocial}, suggesting that interactional toxicity is associated with foreshortened debate and a decreased collective capacity to construct quality Wikipedia pages. 

\begin{figure}
\centering
\includegraphics[width=\linewidth]{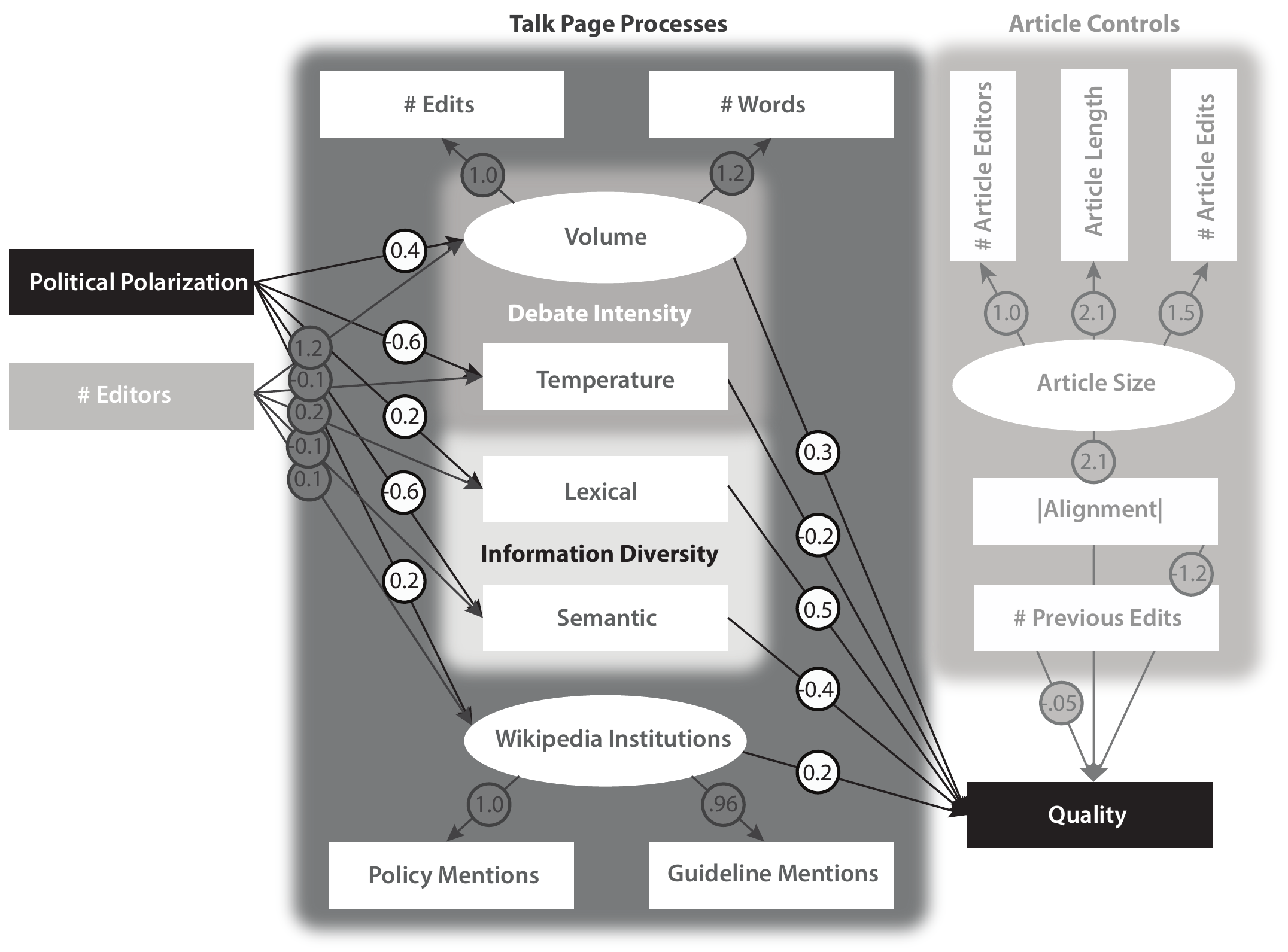}%
\caption{Estimated structural equation model linking \textit{political polarization} (top left) with \textit{article quality} (bottom right) through talk page debate intensity, information diversity and use of Wikipedia policies and guidelines. The right panel includes control variables associated with features of the articles themselves. Rectangles represent measured variables and ovals indicate latent variables. All coefficients are significant at the $p<.0001$ level, agreeing with individual models and bivariate correlations. See Appendix for more details about model and results.}
\label{fig:SEM}
\end{figure}

We present results from a structural equation model in Figure \ref{fig:SEM}, which allowed us to evaluate the combined impact of political polarization on article quality through mechanisms of collaboration. (See Data and Methods and Appendix for additional details.) Compared with politically homogeneous or skewed teams, polarized teams debate \textit{fewer} topics with \textit{more} competing terminology and framings. They engage in \textit{more} debate, which is \textit{less} acrimonious. And they more frequently appeal to Wikipedia policies and guidelines to govern these interactions. 

Mechanisms of polarized collaboration are echoed by editors in their survey responses. One third of respondents indicated awareness of politically motivated conflicts, and two thirds of those described them in detail. Conflicts typically entailed the encounter of biased content (e.g. ``The page read like anti-Russian propaganda''), or having one's own content revised by editors perceived as biased (e.g. ``My neutral edits regarding a particular political group were moved lower in the article to show negative opinions of this group first''). Many such conflicts were resolved through debate. One respondent recalled a conflict over the meaning of the word ``refugee'', which was resolved ``by legal arguments that would convince an impartial observer.'' Another related an intense conflict on a page related to homosexuality, but admitted that as a result ``the article is in a better state.'' Other conflicts were resolved through administrator intervention. One respondent reported editing a page about a far-right politician that other editors would repeatedly vandalize; administrators intervened and protected the page from further edits. Unbalanced political competition, however, where lone editors sought to de-bias articles maintained by politically like-minded communities (e.g., with a perceived ``right wing slant'' or ``anti-Russian bias'') led to more acrimonious conflict that often resulted in editor bans. Editing contested topics required toughness and endurance, which was ameliorated by balanced conflict. It is precisely these engagements that are missing from segregated ``echo chamber'' platforms, and which channel Wikipedia editors' diverse perspectives into articles of superior quality. 

\section*{Discussion}
This study provides the first empirical, real-world evidence that political polarization can lead to productive collaboration. Wikipedia teams comprised by a balance of politically polarized individuals perform better than groups comprised of political partisans and even moderates. Positive effects from polarization are observed in \textit{Political}, \textit{Social Issues}, and even \textit{Science} articles. The intensified effect of political polarization on pages with greater political content suggests that diversity is not universally beneficial, but assists when directly or indirectly relevant to the topics considered. We demonstrate how frequent, intense disagreement within politically polarized teams foments focused debate \cite{Mercier2016-mb} and, as consequence, higher quality edits that are more robust and comprehensive.

The observational nature of this study places constraints on interpreting the relationship between political polarization and quality as a causal one. We observed only the behavior of those editors who voluntarily cooperated with others of contrary politics to produce articles of higher quality, or those who avoided such collaborations and produced lower quality articles. It is possible that these are different kinds of people, and so we cannot rule out the possibility that \textit{randomly} assigned politically polarized teams may not outperform more homogeneous teams. Concerns of extreme self-selection are, however, allayed by Wikipedia's ``encyclopedic monopoly''. As the fifth most visited website in the world with more than 5 million articles on a wide range of topics, Wikipedia represents an effective monopoly of reference attention. Efforts have been made to produce politically skewed alternatives\footnote{https://www.wired.com/story/welcome-to-the-wikipedia-of-the-alt-right/.}, but no viable substitutes exist. More importantly, Wikipedia contains only a single version of an article for a given topic. Consequently, if someone wishes to influence public knowledge on topics such as ``Climate change'' or ``Free market'' through Wikipedia, they must collaborate with existing editors who hold differing views but equal motivation. This is particularly salient for articles on contested topics, and frames a dramatic contrast with segregated ``echo chambers'' in the blogosphere. Previous research on Wikipedia also suggests that cross-party collaboration is the norm rather than the exception \cite{Greenstein2016-ig}.

Politically diverse collaborations are not without costs. One major obstacle to creating well functioning, diverse teams is that such teams produce outputs that may appear worse to the team members themselves \cite{Phillips2009-dt}. Membership in homogeneous teams also \textit{feels} better as participation affirms prior beliefs \cite{Hannah_Nam2013-ti} and shelters contributors from aggressive interaction. Respondents to our survey echoed this sentiment by reporting pervasive displeasure in having to convince obstinate, competing partisans of points that they took to be self-evident. Balanced competition softened the emotional edge of ideological conflict, however, by allowing members to police tone and content with the omnipresent policies and norms of Wikipedia \cite{jemielniak2014common}. Unlike many online settings, when norms and policies break down, powerful moderators may step in and revert edits, lock pages and execute bans. 

Previous research suggests that very high levels of diversity in teams may deteriorate the quality of teamwork. To explore whether political diversity has an upper bound beyond which polarization \textit{hampers} performance, we re-estimated the regression models of quality with a quadratic polarization term. Estimates suggest that quality may eventually decline with increasing polarization, but the optimal level of polarization is above that realized by 95\% of the teams in this study. For the 5\% most polarized teams, there is no statistically significant pattern between polarization and quality. In other words, we do not find evidence that very high levels of political polarization hampers Wikipedia performance.

This study raises the possibility that in crowd-sourcing \textit{contested} knowledge, the most motivated contributors are those with a bias or ``angle'' on the disagreement at hand. Conducting debates on platforms like Wikipedia can require high levels of motivation and patience\footnote{For example, the top editor of Hillary Clinton's Wikipedia page estimated spending 15 hours per week on protecting it from vandals. https://newrepublic.com/article/63288/wiki-woman.}, and neutral users lacking partisan motivation may choose to allocate their time elsewhere. It is plausible that for voluntary crowd-sourcing platforms there exists an \textit{optimal, non-zero} amount of user bias. Platforms that discourage all user bias may thus be inefficient or unsustainable.

Insofar as political diversity can improve the quality of politically relevant crowd-sourced knowledge, it is important to consider whether platforms should intervene to promote or even impose such diversity where missing\footnote{Indeed, platforms like Facebook are moving in precisely this direction: https://newsroom.fb.com/news/2017/10/news-feed-fyi-new-test-to-provide-context-about-articles/.}. Our work suggests that for contested knowledge, platforms should seek not only high \textit{numbers} of experts, but those with balanced, diverse perspectives to construct an environment through which motivated conflicts can be disciplined by enforceable policies and guidelines. Just as institutional designs to promote gender diversity have proven valuable for fairness and performance in a variety of domains \cite{bohnet2016works,dahlerup2012impact,besleygender}, designing for political diversity may become an increasingly important priority. 

\bibliographystyle{model1-num-names}
\bibliography{wikipoli_references.bib}

\pagebreak
\appendix

\section{Descriptive statistics of the dataset} \label{si:stats}

Summary statistics of the three corpora -- politics, social issues, and science articles -- are shown in Table \ref{tab:stats}. We measured the quality of Wikipedia articles algorithmically using a prominent approach that draws on features derived from article content alone and not information about editors or their collaboration patterns \cite{Warncke-Wang2013-ay}. Wikipedia editors have scored hundreds of articles on quality, but human-generated ratings for most of Wikipedia's millions of articles do not exist and necessitate an algorithmic approach. In particular, we used the \texttt{wikiclass} algorithm, developed by Wikimedia research staff \cite{Halfaker2017-cx} and trained on Wikipedia pages scored by active editors for quality using a six-class scale, which ranges from ``Featured Article'' (highest quality) to ``Stub'' (lowest quality).  The \texttt{wikiclass} algorithm predicts the correct quality class in 62.9\% of cases and is off by at most one quality class in 90.7\% of cases \cite{Halfaker2017-cx}. The distribution of estimated quality for each article is shown in Table \ref{tab:quality}. Note that a few articles have no text (e.g., removed or redirected) and hence receive no quality ratings.

\begin{table}[tbhp]
\small
\centering
\begin{threeparttable}
\caption{Summary statistics of Wikipedia data sets}
\label{tab:stats}
\begin{tabular}{lcccc}
Corpus       & \# Articles & Article length & \# Edits per article & \# Editors per article \\
\midrule
Politics\\
\hspace{3mm}Conservative & 10,909      & 9,449 (15,013)   & 177 (808)                 & 80 (294)                  \\
\hspace{3mm}Liberal      & 10,038      & 8,645 (14,280)    & 155 (686)                 & 75 (282)                       \\
Social Issues & 	162,085 & 	13,153 (20,847) & 265 (819) & 122 (337) \\
Science      & 49,995     & 11,193 (17,297)   & 210 (	632)                 & 103 (284)                       \\
All          & 233,027     & -              & -                    & -                      \\
\bottomrule
\end{tabular}
\begin{tablenotes}
\item
\small
Article length, \# Edits per article, and \# Editors per article refer to the averages over all articles in the corresponding corpus. Article length is measured by bytes. Numbers in parentheses are standard deviations.
\end{tablenotes}
\end{threeparttable}
\end{table}

\begin{table}
\small
\centering
\begin{threeparttable}
\caption{Distribution of article quality in each corpus}
\label{tab:quality}
\begin{tabular}{lcccccc}
Corpus & \multicolumn{6}{c}{Quality rating} \\
\midrule
& Stub & Start & C & B & Good Article & Featured Article\\
\cmidrule{2-7}
Politics	&2950&	5009&	3541&	485&	871&	215\\
Social Issues	&30853	&55884	&48292	&14050	&10108&	3814\\
Science	&12192	&16899	&12454	&5323	&1696	&966 \\
\bottomrule
\end{tabular}
\begin{tablenotes}
\small
\item
Number of articles in each quality rating for Politics, Science, and Social Issues pages. The algorithm used to measure article quality is described in Material and Methods: Measurements.
\end{tablenotes}
\end{threeparttable}
\end{table}

\section{Computational measure of political alignment}
\label{sec:computational measure of political alignment}
Many Wikipedia editors carry out general copy editing and curatorial tasks that contribute very few bytes to articles, which lead us to estimate an editor's political alignment, $p$, through a conservative, Bayesian framework. Our approach is designed to avoid data from these numerous curatorial editors to introduce substantial uncertainty into our overall estimation of political preference. We use a ``neutral" prior to down-weight small-sample effects. Mathematically, $p$ is assumed to have a prior distribution $P(p)=Beta(a,b)$ before observing any data, where everyone is assumed to randomly contribute to red or blue articles. Next, the  distribution is updated by Bayes' law with observations $X$ (bytes contributed to conservative articles) and $K$ (bytes contributed to political articles): $p|X\sim Beta(a+X,b+K-X)$. Finally, political alignment is defined as the posterior mean of $p$: $E[p|X]=(X+a)/(K+a+b)$. For presentational purposes, we rescale the alignment scores linearly onto [-1,1] with 0 as neutral point.

In short, political alignment is a scalar between -1 and 1. Casual editors with few contributions will be close to neutral (alignment=0). This alignment measure allows us to quantify the ideological perspective each editor brings to an editing team and, in turn, an edited article. For example, an average alignment score close to 0 for a group of editors suggests balanced participation from both conservatives and liberals in the group. 

To statistically test whether the observed variance of political alignments is greater than expected from chance, we simulate editors who have a given ``budget'' of edits and choose to allocate them to liberal or conservative pages at random. In each simulation, each editor contributes each of her actual edits to either liberal or conservative articles with a probability proportional to the total size of each set of articles. From these simulations, we construct a distribution of the variance of editor alignments and find that the variance of alignments from the simulated editors is statistically lower than the variance observed among real Wikipedia editors. 

\section{Survey measure of political alignment}
We sampled Wikipedia editors having made recent edits ($<1$ month) to at least 1 article in political and/or science articles. Wikipedia required that surveys be individualized, solicitations be made personally on editors' talk pages, not as a batch, and that we engage all individuals solicited in conversation regarding any questions or concerns. This limited the number of surveys that could be performed. The survey first asked whether the respondent resides in the United States. Those responding in the affirmative were then asked ``\textit{Do you generally think of yourself as a Republican, Democrat, Independent or something else?}". The (mutually exclusive) answer choices were 1=\textit{Strong Democrat}, 2=\textit{Not Strong Democrat}, 3=\textit{Independent, Near Democrat}, 4=\textit{Independent}, 5=\textit{Independent, Near Republican}, 6=\textit{Not Strong Republican}, 7=\textit{Other (Please explain)}. Although the computational measure ranges from "Liberal" to "Conservative," the survey question focused instead on specific political party affiliation for concreteness and its long-standing use in survey research. Some respondents chose to instead write-in a political party and, in some cases, mentioned being registered as either Republican or Democrat. 

54\% of respondents (64/118) reported living outside of the United States. These respondents were then asked "Which political party, if any, do you generally identify with?" Some respondents provided parties that they themselves compared with U.S. Democratic or Republican parties. Other responses could not be unambiguously aligned with Democratic or Republican parties and were excluded from calculations. 

Our overall response rate was 24\% (118/500). Figure 4 displays response rates across editors in the range [-1, +1] of (computationally measured) political alignment. Responses were received from each quintile of the alignment distribution, although only 1 of 27 solicited users with alignment in [0.2, 0.6) replied. 

\begin{figure}[ht]
\centering
\includegraphics[width=0.5\textwidth]{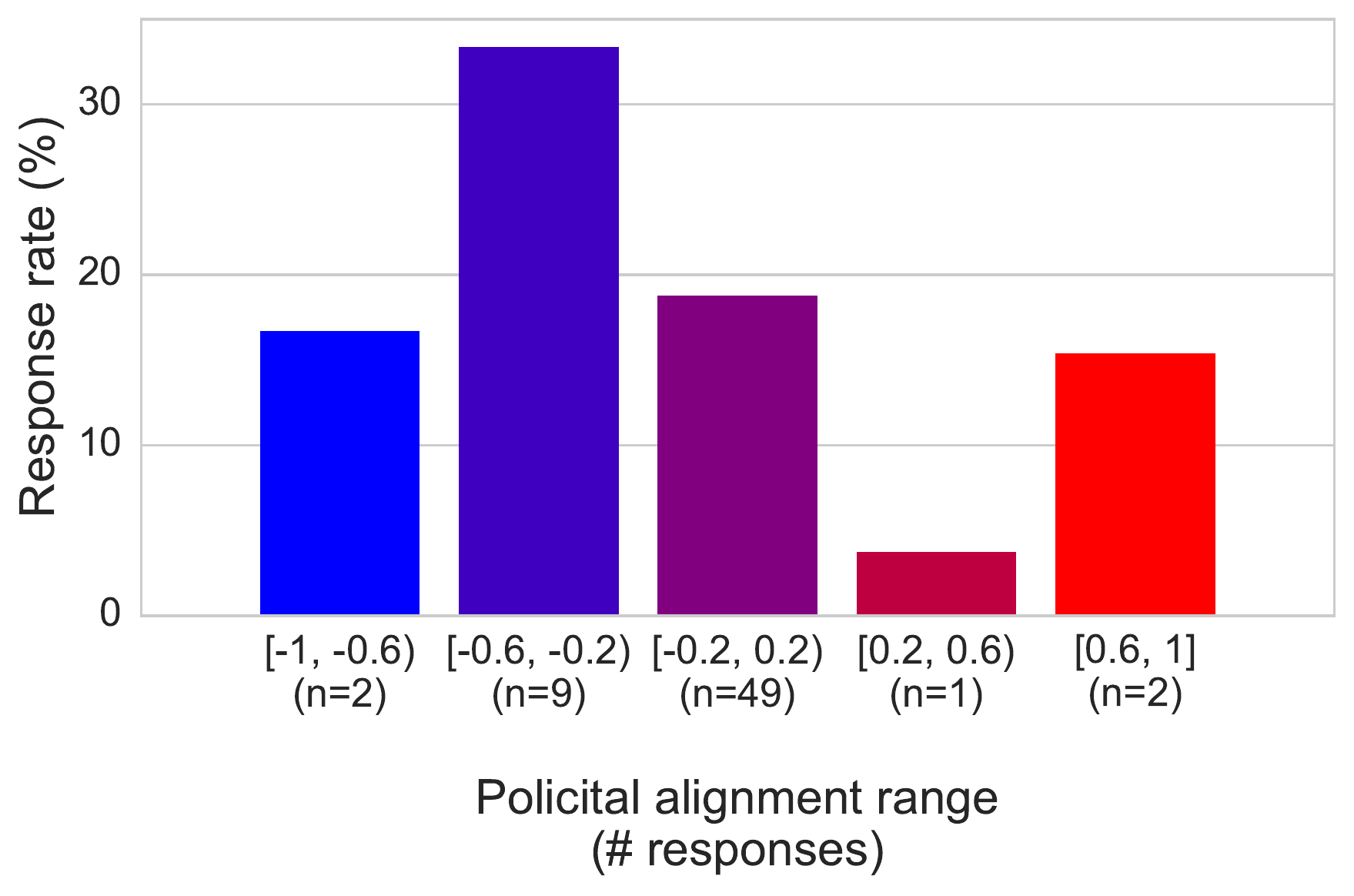}
\caption{Response rate by (computationally measured) political alignment. Each bar is the response rate achieved for editors whose alignments fall in the range specified at the bottom of the bar. Not shown are editors who edited Science articles only, and therefore lack a computationally measured political alignment. Note that not all responses provided useful political data and so not all are represented in Table 3.}
\end{figure} 

Table \ref{tab:surveyvscomputational} reports three correlations between computational and survey measures of political alignments, where different sets of survey responses are used. In each case respondents choosing ``Independent'' were not used in correlations below, because ``Independent'' was assumed to reflect political alignments that do not clearly align on a Conservative-Liberal spectrum. 

\begin{adjustbox}{max width=\linewidth}
\begin{threeparttable}[width=\linewidth]
\caption{Correlations between survey \& computational measures}
\label{tab:surveyvscomputational}
\begin{tabular}[]{rccccc}
 & Editors' location & ``Other party'' re-coding & \# Responses & Corr. coeff. & 2-tailed \textit{t}-test (1-tailed) \\
\midrule
(a)& 	US-based 	& N 	& 21 	& 0.31 	&  p=0.167 (0.083) \\
(b)&	US-based 	& Y 	& 28 	& 0.35 	& p=0.071 (0.036) \\
(c)&	All      	& Y	 & 45 	& 0.31 	& p=0.036 (0.018) \\
\bottomrule
\end{tabular}
\begin{tablenotes}
\item
\small
\textit{Note:} Correlations between self-reported political identification from survey respondents and computationally measured political alignments. Editors' locations are self-reported. Row (a) uses raw survey responses (no recoding). Row (b) adds 7 responses of \textit{Other (Please explain)}, recoded to either \textit{Independent, Near Democrat} or \textit{Independent, Near Republican}. Row (c) adds 17 respondents from non-US locations whose responses could be recoded straightforwardly as in (b).  
\end{tablenotes}
\end{threeparttable}
\end{adjustbox}

\bigskip
Comparison (a) is strictest, using raw responses from US-based editors only. The point-estimate of the correlation, based on n=21 responses, is 0.31 (2-tailed \textit{t}-test \textit{p}=0.167, 1-tailed \textit{t}-test \textit{p}=0.083). 
Comparison (b) adds to (a) responses from US-based editors that selected ``Other'' for political party identification and provided comments that could be recoded to the either ``Near Republican'' or ``Near Democrat'' unambiguously. Examples of the 7 recoded responses include ``In between Libertarian and Republican''$\to$\textit{Near Republican} and ``Social Democrat''$\to$\textit{Near Democrat}. The estimated correlation is 0.35 (n=28, 2-tailed \textit{t}-test \textit{p}=0.071, 1-tailed \textit{t}-test \textit{p}=0.036). 

Comparison (c) adds to (b) respondents who identified their location as outside the US. Many such respondents identified themselves with US-based political parties. We recoded these parties to either ``Independent, Near Republican'' or ``Independent, Near Democrat.'' Examples of recoded responses include ``Labour (UK)''$\to$\textit{Independent, Near Democrat} and ``Left / progressive / social democracy''$\to$\textit{Independent, Near Democrat}. 
The correlation is within the range of comparisons (a-b), but with the higher number of responses (n=47), it is statistically significant at the 0.05 level (p=0.02). 

All three comparisons show that our computational measure of political alignment correlates moderately well (0.31-0.35) with self-reported political identification, suggesting that it is a noisy but valid measurement of editors' political alignments. We focus on comparison (b) because it includes more respondents than (a) but avoids the potentially subjective recoding of non-U.S. parties to the U.S. political spectrum. 

\section{Statistical Analysis}
\subsection{Bivariate Correlation}
\begin{figure}
\centering
\includegraphics[width=1.1\textwidth]{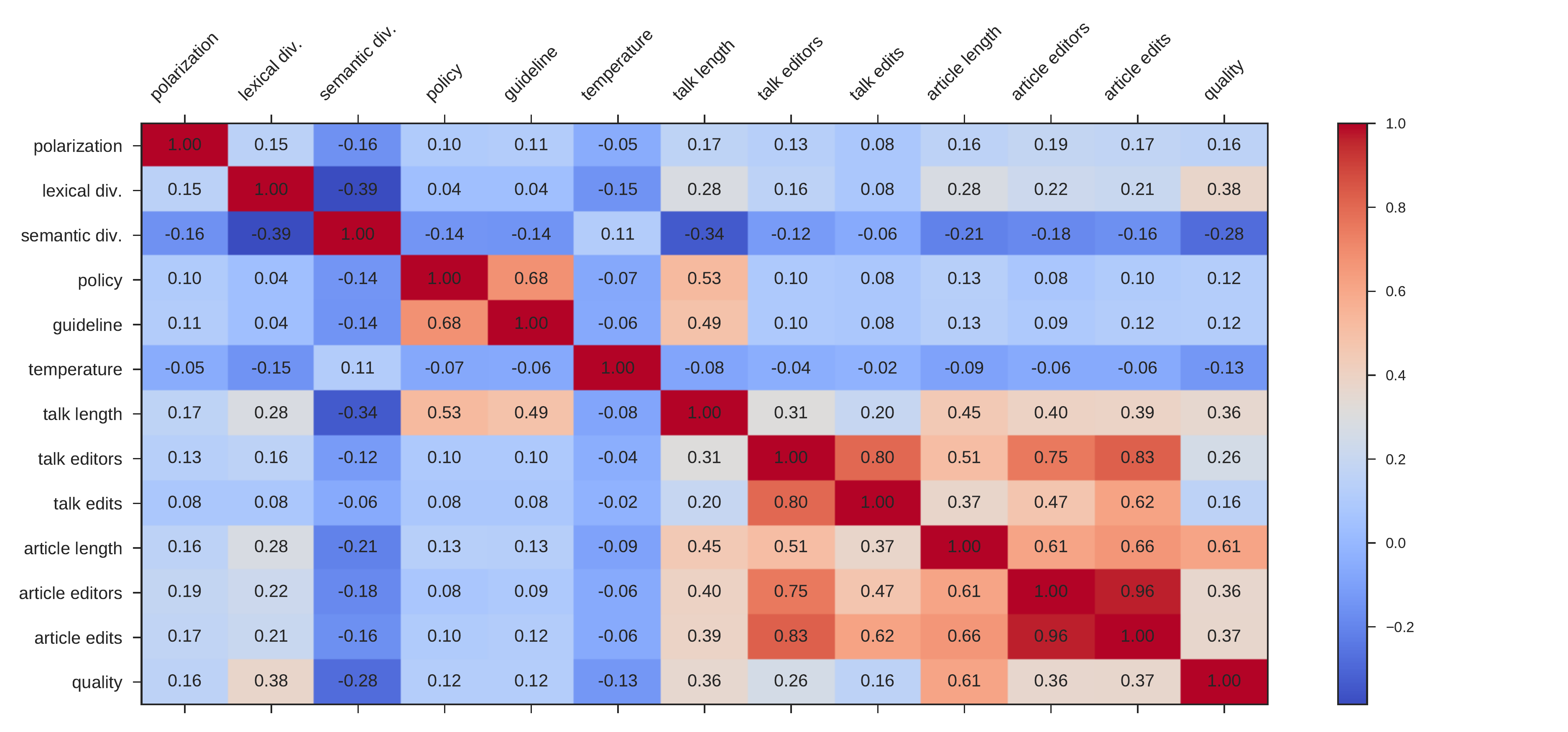}
\caption{Pearson correlation between every pair of variables.}
\label{fig:correlation}
\end{figure}

We explore relationships between variables considered in the study by calculating the Pearson correlation between every pair; the table of correlations is shown in Figure \ref{fig:correlation}.

From the table we can see that polarization (i.e., variance of political alignments) is positively correlated with quality. Besides, polarization is correlated with all other variables in directions  consistent with hypothesized mechanisms of collaboration: 
\begin{itemize}
\item positively with lexical diversity, and negatively with semantic diversity, suggesting that polarization will focus debate on fewer contested, politically relevant topics, but frame them in competing ways.
\item positively with all talk page and article activities such as page length, number of edits, and number of editors, suggesting polarization increases debate volume.
\item negatively with talk page temperature, suggesting that polarization, resulted from balanced engagement, polices and decreases emotional aggression and violent debate.
\item positively with policy and guideline mentions, suggesting increased use of Wikipedia institutions among polarized teams.
\end{itemize}
The table also reveals that most variables are substantially and significantly correlated with each other. Correlation between polarization and other variables could be caused by confounding factors. Therefore, we carried out two further statistical analyses that estimate the conditional effects of polarization, holding other confounding variables constant. First, we estimated a number of individual regressions, linking polarization to quality through all proposed mechanisms. Then we estimated a structural equation model, which evaluates all individual effects simultaneously, enabling us to compare their relative strength. 

\subsection{Regression Analysis}
To assess the mechanisms of polarized collaboration, we estimated multiple linear regression models with polarization as the main predictor. Specifically, we tested how polarization (i.e., variance of alignments) affected, respectively, talk page volume (i.e., number of talk page edits, talk page length), semantic and lexical diversity, policy and guideline mentions, and talk page ``temperature'', controlling for several Wikipedia talk and article page features. Models and regression results are shown in Table \ref{tab:regression}.

All models control for number of talk page editors to make sure that the effects are not simply caused by number of people involved. Talk and article page lengths are also accounted for whenever possible, so the effects of polarization are not only due to the sheer amount of edits and words. Lexical and semantic diversity are normalized by page length so it is not necessary to include page length in those models; instead, we control for lexical and semantic diversities of the articles to account for heterogeneous article content.

Note that we do not include every variable in every model because of substantial collinearity between some variables (see the correlation table above). Simpler models are more interpretable and preferable from a statistical point of view. For example, number of editors, number of edits and page length are highly correlated with each other; when all are included in a single model, it is hard to explain the direction of their effects (e.g., in the 2nd model in Table \ref{tab:regression}, number of editors shows a negative effect on page length when number of edits is present). These encourages us to create factors from highly correlated and conceptually similar variables. We do this, and simultaneously evaluate the combined impact of polarization on article quality through estimation of a structural equation model as described below. 

\begin{adjustbox}{max width=\linewidth}
\begin{threeparttable}
\caption{Regression results from 7 models that predict qualities of Wikipedia talk page deliberation with polarization, controlling for relevant talk and article page features.}
\label{tab:regression}
\begin{tabular}{rccccccc}
\multicolumn{1}{c}{\textit{}} & \multicolumn{7}{c}{\textbf{Dependent variable}} \\
\textbf{Independent variable} & \textit{\# talk edits} & \textit{talk page length} & \textit{lexical diversity} & \textit{semantic diversity} & \textit{temperature} & \textit{policy}      & \textit{guideline}   \\ 
\midrule
polarization                  & 0.10                & 1.28                   & 1.43                    & -0.44                    & -0.17             & 0.06              & 0.09              \\
\# talk editors               & 1.03                & -0.09                  & -0.09                   & 0.15                     & 0.25              & -0.28             & -0.25             \\
\# talk edits                 &                        & 1.13                   & 1.39                    & -0.21                    & -0.03             & 0.29              & 0.26              \\
talk page length              & 0.09                &                           &                            &                             & -0.21             & 0.05              & 0.05              \\
article length                & 0.05                & 0.24                   &                            &                             & -0.09             & -0.01             & -0.004            \\
article lexical d.            &                        &                           & 0.10                    &                             &                      &                      &                      \\
article semantic  d.          &                        &                           &                            & 0.34                     &                      &                      &                      \\
\multicolumn{1}{l}{}          & \multicolumn{1}{l}{}   & \multicolumn{1}{l}{}      & \multicolumn{1}{l}{}       & \multicolumn{1}{l}{}        & \multicolumn{1}{l}{} & \multicolumn{1}{l}{} & \multicolumn{1}{l}{} \\
\textit{$R^2$}                & 0.95                   & 0.71                      & 0.68                       & 0.21                        & 0.17                 & 0.27                 & 0.26                 \\ 
\bottomrule
\end{tabular}
\begin{tablenotes}
\item
\small
Note: All variable coefficients are statistically significant at the $p<0.001$ level. Both dependent and independent variables are log-transformed so that distributions are made approximately Gaussian.  
\end{tablenotes}
\end{threeparttable}
\end{adjustbox}

\subsection{Structure Equation Modeling}
The structural equation model we designed is detailed in Figure 4 of the main text, and estimated by the R package ``lavaan'' \cite{JSSv048i02}. Detailed specifications and estimation results are as follows.

\begin{itemize}
\item Latent variables: debate volume, institution, article activity.
\begin{eqnarray}
\mbox{debate volume} &\sim & \mbox{number of talk edits} + 1.15 \mbox{ talk page length} \nonumber \\
\mbox{institution} &\sim & \mbox{policy mentions} + 0.96 \mbox{ guideline mentions} \nonumber \\
\mbox{article activity} &\sim & \mbox{number of article editors} + 2.1 \mbox{ article length} + 1.48 \mbox{ number of article edits} \nonumber
\end{eqnarray}

\item Regressions
\begin{eqnarray}
\mbox{quality} &\sim & -1.19~|\mbox{average alignment}| -0.05 \mbox{ editing experience} + \nonumber\\
&& 2.14 \mbox{ article activity} +  0.33 \mbox{ debate volume} + \nonumber\\
&& 0.50 \mbox{ lexical diversity} -0.35\mbox{ semantic diversity} + \nonumber\\
&& 0.18\mbox{ institution} -0.23\mbox{ debate temperature} \nonumber \\
\mbox{debate volume} &\sim & 0.38 \mbox{ polarization} + 1.19 \mbox{ number of talk editors} \nonumber \\
\mbox{lexical diversity} &\sim & 0.20 \mbox{ polarization} + 0.17 \mbox{ number of talk editors} \nonumber \\
\mbox{semantic diversity} &\sim & -0.65 \mbox{ polarization}  -0.1 \mbox{ number of talk editors} \nonumber \\
\mbox{institution} &\sim & 0.20 \mbox{ polarization} + 0.14 \mbox{ number of talk editors} \nonumber \\
\mbox{debate temperature} &\sim & -0.80 \mbox{ polarization} -0.12 \mbox{ number of talk editors} \nonumber
\end{eqnarray}

\item Effects of polarization on quality through the following paths:
\begin{eqnarray}
\mbox{polarization} \rightarrow & \mbox{debate volume} & \rightarrow \mbox{ quality}: 0.125 \nonumber \\
\mbox{polarization} \rightarrow & \mbox{lexical diversity} & \rightarrow \mbox{ quality}: 0.098 \nonumber \\
\mbox{polarization} \rightarrow & \mbox{semantic diversity} & \rightarrow \mbox{ quality}: 0.230 \nonumber \\
\mbox{polarization} \rightarrow & \mbox{institution} & \rightarrow \mbox{ quality}: 0.035 \nonumber \\
\mbox{polarization} \rightarrow & \mbox{talk temperature} & \rightarrow \mbox{ quality}: 0.136 \nonumber
\end{eqnarray}

\item Combined effect of polarization on quality through all the paths: 0.624.
\end{itemize}

The model is well specified and the estimation procedure converged quickly (125 iterations). All estimated parameters in the model are significant at $p<0.001$, agreeing with the other statistical analyses. 

Because the sample size is very large (205,749 observations), a $\chi^2$ test cannot be used to evaluate this model. (In fact, the $p$-value of a $\chi^2$ is approximately 0.) The CFI and NNFI indexes for the model are 0.78 and 0.71, respectively. We do not expect the indexes to be very high because the model is designed to test the effects of polarization through various mechanisms rather than fitting all covariances in the data. For example, the correlation table reveals that article activity is highly correlated with debate volume. If we add a regression of article activity on volume to the current model, the CFI index boosts to 0.89, but the interpretation becomes less clear as a new path is introduced through article activity.  

\end{document}